\documentclass[letterpaper,amsmath,amssymb,floatfix,aps,superscriptaddress,nofootinbib]{revtex4} 

\usepackage{graphicx}% Include figure files
\usepackage{dcolumn}% Align table columns on decimal point
\usepackage{bm}% bold math
\usepackage{epsfig,color}

\newcommand {\vct}[1] {\mathbf  #1}

\def\ul#1#2{\textstyle{\frac#1#2}}
\def\rmd{{\mathrm{d}}}
\def\rmi{{\mathrm{i}}}

\def\rme{{\mathrm{e}}}

\begin{document}

\title{Exotic Electrostatics: Unusual Features of Electrostatic Interactions between Macroions}

\author{Ali Naji}
\affiliation{Department of Applied Mathematics and Theoretical Physics, 
%Centre for Mathematical Sciences, 
University of Cambridge, Cambridge CB3 0WA, United Kingdom}
\author{Matej Kandu\v c}
\affiliation{Department of Theoretical Physics,
J. Stefan Institute, SI-1000 Ljubljana, Slovenia}
\author{ Roland R. Netz}
\affiliation{Department of Physics, Technical University of Munich, James Franck Strasse, 
D-85748 Garching, Germany}
\author{Rudolf Podgornik}
\affiliation{Department of Theoretical Physics,
J. Stefan Institute, SI-1000 Ljubljana, Slovenia, and 
Department of Physics, Faculty of Mathematics and Physics and Institute of Biophysics, Medical Faculty, 
University of Ljubljana, SI-1000 Ljubljana, Slovenia}

\begin{abstract}
We present an overview of our understanding of electrostatic interactions between charged
macromolecular surfaces mediated by mobile counter- and coions. The 
dichotomy between the {\em weak} and the {\em strong coupling} regimes is described in detail and
the way they engender repulsive and attractive interactions between nominally equally 
charged macroions. We also introduce the concept of {\em dressed counterions} in the
case of many-component Coulomb fluids that are partially weakly and partially strongly coupled to 
local electrostatic fields leading to non-monotonic interactions between equally 
charged macroions. The effect of quenched surface charge disorder on the counterion-mediated electrostatic interactions is analyzed within the same conceptual framework and shown to lead to unexpected and extraordinary electrostatic interactions between randomly charged surfaces with equal mean surface charge densities or even between effectively 
neutral macroion surfaces. As a result, these recent developments challenge some cherished notions of pop culture.
\end{abstract}

\maketitle

%%%%%%%%%%%%%%%%%%%%%%%%%%%%%%%%%%%%%%%%%%%%%%%%%%%%%%%%%%%%%%%%%%%%
\section{Introduction}
\label{sec:intro}

The nature of electrostatic interactions between charges started to be studied intensively in the second half of the 18th century \cite{whittaker}. Benjamin Franklin first inferred from an observation that surprisingly there is no force on a charge inside a charged sphere, an observation later repeated by Joseph Priestley (1767). John Robison (1769) determined that the electrostatic force falls off with (almost) the second power of separation between charges. Based on the work of Charles A. de Coulomb (1777) who is a co-inventor of the torsion balance for measuring the force of magnetic and electrical attraction,  Henry Cavendish measured directly the interactions between charges (1779) but did not publish his results. They were eventually published by William Thomson (Lord Kelvin) one hundred years after the original discovery (1879). From hard science electrostatic interaction penetrated the pop culture in general and today everybody knows that {\em opposites attract} and {\em likes repel!} 

The exact form of the electrostatic interaction is since known to be given by the Coulomb's law which states that the interaction potential between two charges $e_1$ and $e_2$ {\em in vacuo} located at ${\bf r}$ and ${\bf r}'$ respectively, can be written in the standard form (in SI units) as
\begin{equation}
V({\bf r}, {\bf r}') = \frac{e_1 e_2}{4 \pi \varepsilon_0 \vert{\bf r} - {\bf r}'\vert},
\label{eq1.1}
\end{equation}
were $\varepsilon_0$ is the permittivity of vacuum. Electrostatic interaction is the fundamental interaction in molecular world giving rise to short range intramolecular bonds as well as longer ranged interactions between molecules and their aggregates \cite{French-RMP}. While atomic bonds are obtained by combining the Coulomb potential with the quantum of action, longer ranged colloidal and nanoscale electrostatic interactions are obtained by combining the Coulomb potential with the thermal energy $k_{\mathrm{B}}T$. 

Almost exactly a hundred years ago Gouy \cite{Gouy} and Chapman \cite{Chapman} were the first to combine thermal energy and Coulomb interactions into a statistical theory of Coulomb fluids basing their approach on what became later known as the {\em Poisson-Boltzmann (PB) equation}. {\em Coulomb fluid} in general is an assembly of (variously) charged particles in thermal equilibrium. This {\em mean-field} approach to Coulomb fluids was gradually elaborated  in uncountably many ways starting from the seminal work of Debye and H\"uckel \cite{DH} and finally codified as a cornerstone of the fundamental DLVO theory of colloidal interactions by Derjaguin and Landau \cite{DL} as well as Verwey and Overbeek \cite{VO}, where it figures as the repulsive electrostatic part of the total  disjoining pressure acting between macromolecular bodies \cite{Israelachvili,Hunter}. The attractive component in this case is provided solely by the van der Waals interactions that have their origin in the quantum and thermal fluctuations of electromagnetic fields \cite{adrianbook}.

One could claim that modern formulation of statistical mechanics of Coulomb fluids starts with the work of Edwards and Lenard \cite{Edwards} where the grand-canonical partition function of a Coulomb fluid was written in the form of a functional integral over fluctuating electrostatic fields. While Podgornik and \v Zek\v s \cite{podgornik} realized that the collective description based on the mean-field PB theory arises from the {\em saddle-point approximation} to this functional integral in the case of a {\em weak coupling regime} corresponding, for instance, to small external charges, Netz showed \cite{Netz01, AndreNetz}  that another approximation valid in the regime of large external charges leads to a completely different fixed point, formulated in terms of a single-particle description of the same system. This so-called {\em strong coupling regime} has been the focus of various theoretical studies over the past decade \cite{hoda_review,Naji_PhysicaA,Rouzina96, Shklovs02, Levin}.  The two approximations were shown to correspond to extremal values of a single electrostatic coupling parameter and to consistently bracket all available simulation results on statistical properties of non-homogeneous Coulomb fluids \cite{hoda_review,Naji_PhysicaA}, i.e., Coulomb fluids confined between charged boundaries. 

In what follows we shall describe the main consequences of this {\em weak-strong coupling dichotomy} especially as they transpire
in the case of interactions between charged (planar) macromolecular surfaces across a Coulomb fluid comprising mobile counter- and coions.
We shall see that the counterion-mediated electrostatic interactions are strictly repulsive in symmetric
external charge configurations for the case of weak coupling but can turn attractive for strongly coupled surfaces. In the case of many-component systems, composed typically of a weakly coupled univalent salt and strongly coupled polyvalent counterions, these interactions show a subtle interplay between repulsions and attractions for nominally equally charged surfaces. On the other hand, the presence of quenched charge disorder on bounding surfaces can lead to pronounced electrostatic attraction. 
These exotic features only arise in special cases of strongly coupled systems  and can not be rationalized within the standard mean-field PB approach.

%%%%%%%%%%%%%%%%%%%%%%%%%%%%%%%%%%%%%%%%%%%%%%%%%%%%%%%%%%%%%%%%%%%%
\section{Scenery}
\label{sec:scenery}
 
Electric charges and electrostatic interactions are ubiquitous  in 
soft-matter and biological systems \cite{holm,Andelman}.  {\em Soft materials} are typically composed of {\em macromolecules} such as 
polymers, colloids and proteins which often acquire surface charges when dissolved
in a polar solvent like water. This is usually due to 
dissociation of surface chemical groups, which leaves permanent charges on macromolecular  surfaces  
and releases oppositely charged microscopic {\em counterions} into the solution. 
Soft materials are easily deformed or re-arranged by interaction potentials comparable in magnitude to
thermal energy. It thus becomes clear that electrostatic interactions, that  
are typically long-ranged and strong, constitute a prominent factor in
determining the behavior and properties of soft materials. 
This  makes charged materials central to many technological 
applications  and on the other hand, a challenging subject for fundamental research in inter-disciplinary sciences \cite{French-RMP}. In what follows, we briefly review a few examples to demonstrate the diversity of phenomena
associated with charged soft-matter systems.

%%%%%%%%%%%%%%%%%%%%%%%%%%%%%%%
\subsection{Colloids, polymers and membranes: The mesoscopic scale}

Colloids are abundant in nature and industry: smoke, fog, milk, paint and ink are only 
a few examples of colloidal systems. They comprise tiny
solid or liquid particles that are suspended in another medium such as air or another liquid. 
An important factor, which makes colloidal solutions in many ways different from molecular or simple electrolyte solutions  
(such as sugar or salt solution), is the large asymmetry in size and mass between the 
colloidal particles and solvent molecules (or microscopic ions): colloids are {\em mesoscopic} or even {\em nanoscopic} particles with sizes in the range of a few nanometers to microns
that are indeed made of many atoms, but not yet sufficiently many to make 
them behave like macroscopic bodies. 

In colloidal dispersions, the total area that is in contact with solvent is tremendously large: for nanometer-sized colloids, 
nearly half of the atoms are at the surface whereas for macroscopic bodies this ratio tends to zero. 
This is even more true for extended quasi-two-dimensional macromolecular aggregates such 
as lipid membranes and surfactant-covered interfaces  that can carry a substantial amount  of charged molecular moieties \cite{Andelman}. 
Therefore, contrary to the typical situation in the macroscopic world, 
the physics of meso- and nanoscopic particles are dominated by  ``surface" properties and interactions \cite{VO,Israelachvili,Hunter}.

Another relevant mesoscopic or {\em macromolecular} system are polymers (with everyday-life examples like
chewing gum, dough or egg white),  in which many repeating
subunits (monomers) are chemically connected to form a flexible chain. 
Flexible polymers are distinguished by their many degrees 
of freedom associated with conformational rearrangements of monomers that are easily excited 
by thermal energy at room temperature 
leading to a diverse phase behavior spanning extended and strongly entangled polymers such as DNA in solution, all the way to
collapsed polymer chains organizing into compact globular states as in the case of proteins. 
Depending on their chemical structure, polymer chains can
have a large mechanical stiffness as well, behaving like rigid rods at small length scales, or can be substantially charged
giving rise to {\em polyelectrolytes}, in both cases playing an important role in biological processes occurring in the cell \cite{MBOC}.
%smaller than a characteristic persistence length (Figure \ref{fig:polymer_colloid}a). Two famous examples  
%of stiff polymers are provided by Nature: Actin filaments and microtubules that are known by their
%important 

%-------------------------
\begin{figure}[t]
\begin{center}
\includegraphics[angle=0,width=8.0cm]{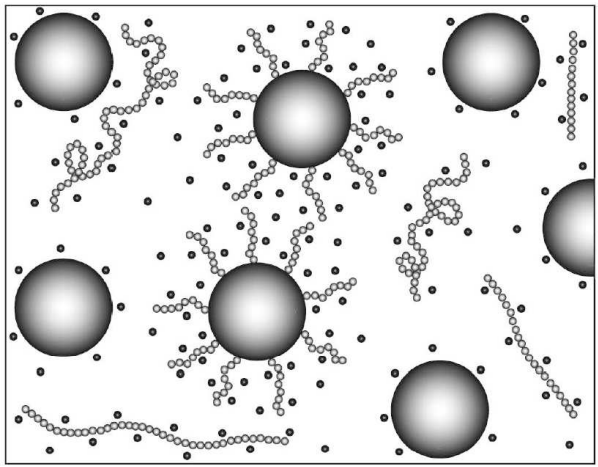}
\caption{\label{fig:polymer_colloid} 
Schematic view of a charged solution consisting of charged colloids (dark spheres), 
charged polymer chains and microscopic neutralizing counterions (small black spheres).
Colloids may be covered by charged polymer brushes, which 
%are typically highly swollen in aqueous solutions and 
generate an additional repulsive interaction between them.
}
\end{center}
\end{figure}
%---------------------
 
%%%%%%%%%%%%%%%%%%%%%%%%%%%%%%%
\subsection{Charges: from industry to biology}

In the mesoscopic world, only electromagnetic interactions that emerge in a 
variety of forms, are important; they are capable of overcoming thermal fluctuations--which are characterized 
by an energy scale of about 0.025$e$V at room temperature \cite{Israelachvili}--and  thus enable formation of stable {\em condensed} phases for soft complex materials.   

In general, colloids dissolved in an aqueous solvent  attract each other due to ubiquitous van der Waals dispersion forces that result from induced  charges 
%(via spontaneous polarization of atoms) 
on their surface \cite{French-RMP}.   As a result, colloidal particles  tend to form large aggregates that typically sediment and destroy the dispersion.  In many applications (for example in food emulsions such as  milk), however, 
stability of a colloidal dispersion 
%against coagulation or flocculation (that is sedimentation of large aggregates in the solution) 
is a desirable property. One way to stabilize dispersions against aggregation is to generate
long-range repulsive interactions between colloidal particles by charging 
their surfaces, which leads to the DLVO mechanism for the stability of  
colloidal dispersions \cite{VO,DL}. Another method of stabilization is to end-graft polymer chains (or {\em polymer brushes}) 
to the particle surfaces \cite{Napper}. For this task, charged polymers are ideal since they 
swell substantially in aqueous solutions and inhibit close contact 
between colloids (see Fig. \ref{fig:polymer_colloid}). 
This latter mechanism has the advantage that it is highly stable against the addition
of electrolyte or salt ions \cite{PIN91}. 
%But since dense charged polymer brushes trap a large amount of oppositely charged ions (counterions) inside, the structure of the  brush layer, and thus the repulsive force generated between colloids,  remains highly  insensitive to the amount of additional salt, which is a common ingredient of colloidal solutions. 

Charged polymers, or {\em polyelectrolytes}, play a significant role 
in the production of cheap, non-toxic and environmentally friendly
materials \cite{Polyelec,Oosawa}.  In contrast to water-insoluble hydrocarbon chains, 
polyelectrolytes typically show high affinity for water and heavy metal ions, 
which makes them useful in applications such as  super-absorbing diapers, waste water purifiers and washing 
detergents and their additives. 

 %-------------------------
%\begin{figure}[t]
%\begin{center}
%\includegraphics[angle=0,width=10.0cm]{DNAfig.eps}
%\caption{\label{fig:DNA} 
%Left: Single DNA molecule of the T2 bacteriophage (about  50 $\mu$m long) is tightly packed  
%into the viral capsid (shown in the middle). 
%When exposed to distilled water, the capsid shell is raptured (osmotic shock) due to the excess internal osmotic pressure, 
%and DNA spills out. Adapted from Ref. \cite{Kleinschmidt}~.
%Right: Three now-empty T5 bacteriophages, shown on top left, 
%have injected their DNA into a liposome. The remaining bacteriophage, bottom left, 
%is about to inject its DNA \cite{Lambert}. The subtle electrostatic effects inside the liposome due to 50 mM  of trivalent spermidine chloride have caused the three viral DNA molecules (of total length of about 124 $\mu$m) 
%121,400bp each amounting to 41,276 nm
%to form a tightly wound toroidal condensate (the capsid
%size is about 70 nm). Adapted from Ref. \cite{PhysToday}~.
%The medium contains 50 mM of spermidine chloride that yields a sufficient concentration of  trivalent cations to condense the DNA. 
%}
%\end{center}
%\end{figure}
%---------------------

In biology electrostatic effects between charged polymers such as DNA and RNA are ubiquitous \cite{holm}. 
DNA, for instance, is a long biomolecule with a total length  of about two meters in human cells, bearing one elementary charge  per 1.7\AA, which for human DNA adds up to $10^{10}$ elementary charges overall!
Yet the DNA is densely packed inside the cell nucleus with a diameter of
less than a few microns. In eucaryotic cells, this storage process involves a hierarchical structure on the lowest level of which short segments of DNA are tightly wrapped around positively charged histone protein complexes 
of a few nanometers in diameter \cite{MBOC}.  This protein-DNA complexation is believed to be governed by electrostatic interactions \cite{Kunz,Hoda,hoda_review}. Electrostatic effects also play a  key role  in complexes of DNA with cationic lipids \cite{Raedler97,Raedler98,Raedler98b}, which are promising synthetically based non-viral carriers of DNA for gene therapy \cite{Crystal}. 

Another example (which is closely related to the results presented later in this chapter)  
is the DNA condensation \cite{holm,PhysToday,Bloom}, in which electrostatic effects enter in a counter-intuitive fashion: here like-charged segments of DNA strongly attract each other! 
In the {\em in vitro} experiments \cite{Lambert}, the condensation of DNA is realized using bacteriophages, which consist of a rigid shell (the capsid) that accommodates a single molecule of viral DNA. These viruses can inject their DNA into a cell or a lipid vesicle. 
% (see Figure \ref{fig:DNA}). 
As a result, large lengths of DNA (up to a hundred microns) can be fitted and condensed into 
a tightly packed, circumferentially wound torus with a diameter of about a hundred nanometers. 
% (see Figure \ref{fig:DNA}).  
This packaging process, which works against the Coulomb self-repulsion and the conformational entropy of the DNA chain, 
is  facilitated and depends upon the presence of  high-valency counterions in the medium. 
Similarly, other highly charged polymers, such as negatively charged F-actin and microtubules 
%, a principal structural protein in cells and muscle tissues, 
can aggregate into closely packed rod-like bundles when small amounts of 
polyvalent cations are added to the solution \cite{Tang96,Tang97,Needleman}.  
It turns out that, in general, when particles are strongly charged, the role of electrostatic interactions 
dramatically changes \cite{hoda_review}: here electrostatic interactions  themselves can trigger the destabilization of charged solutions by mediating attractive like-charge interactions!

%%%%%%%%%%%%%%%%%%%%%%%%%%%%%%%
\subsection{Theoretical challenge and coarse-grained models}

From a theoretical point of view, charged systems pose a many-body problem: 
{\em macroions}, such as charged colloids and polymers, and other charged macromolecular surfaces, such as lipid membranes
and surfactant layers, are always surrounded by 
counterions, and also in general by coions. These particles form loosely bound ionic clouds around macroions and 
tend to screen their charges.  In particular, counterions that are attracted towards macromolecular surfaces,   
predominantly determine the static and dynamic  properties of macromolecular solutions. Understanding the interactions between macromolecules across an ionic medium thus requires an understanding of the counterionic clouds first.

In the most common theoretical approaches known also as {\em primitive models}, the molecular nature 
of the solvent is neglected and is represented by a continuum dielectric medium. 
In reality, the solvent structure is locally perturbed around particles, which can give rise to
additional short-ranged solvent-induced interactions \cite{Israelachvili,benyaakov,Burak_solvent}. 
On the other  hand,  the microscopic features of the macroions are taken into account
using coarse-grained models that incorporate only a few effective parameters such as an effective surface 
charge density.  In most cases, the specific effects associated with ions \cite{benyaakov} as well as the image charge effects due to dielectric inhomogeneities are also neglected. These models therefore  represent a {\em crude simplification} of reality, yet given those simplifications, they can still lead to a systematic and clear understanding of electrostatic effects. 

Here we shall first begin by adopting such a simple model for the interaction
between charged macromolecular surfaces in the presence of {\em counterions only}, but 
then examine the effects due to the additional salt \cite{SCdressed} and the heterogeneous or {\em disordered} distribution of surface charges \cite{ali-rudi,rudiali,disorder-PRL, partial} in more detail. For simplicity, we shall also focus only on the case of {\em planar} charged surfaces, appropriate for the case of charged membranes, solid substrates, or large colloids. Other factors such as non-planar geometry of charged surfaces \cite{Naji-cylinders,Naji_CCT,Matej-cyl}, image charges \cite{jho-prl, kanduc,rudiali,Matej-cyl},  dissimilar surfaces \cite{asim} or multipolar structure of counterions \cite{multipoles} have been analyzed within the same context.

%%%%%%%%%%%%%%%%%%%%%%%%%%%%%%%%%%%%%%%%%%%%%%%%%%%%%%%%%%%%%%%%%%%%
 \section{Length scales in a classical charged system}
\label{sec:scales}

Consider a system of fixed charged objects  with uniform surface charge density 
$-\sigma_{\mathrm{s}} e_0$ (with $e_0$ being the elementary charge) that are surrounded by 
their neutralizing counterions of charge valency $+q$ in a solvent of
dielectric constant $\varepsilon$ at temperature $T$.\footnote{We conventionally assume that macroions are negatively charged and counterions are positively charged, thus $\sigma_{\mathrm{s}}$ and $q$ are both positive by definition.}
%The solvent is assumed to be a continuum medium of  dielectric constant, $\varepsilon$, independent of the temperature, $T$.

One of the basic characteristic length scales in a charged system 
is the {\em Bjerrum length}  \cite{Bjerrum}
\begin{equation}
  \ell_{\mathrm{B}}= {e_0^2}/({4\pi \varepsilon \varepsilon_0 k_{\mathrm{B}}T}), 
\label{eq:Bj}
\end{equation}
which is set by the ratio between  
the thermal energy, $k_{\mathrm{B}}T$, and  the Coulomb interaction energy between two elementary charges at separation $r$, i.e.,   
$V/(k_{\mathrm{B}}T)=\ell_{\mathrm{B}}/r$. The Bjerrum length  thus 
measures the distance at which two elementary charges interact with an energy equal to $k_{\mathrm{B}}T$. In water and at room temperature ($\varepsilon=80$), one has  $\ell_{\mathrm{B}}\simeq 7.1$\AA. For counterions of charge valency $+q$, the Bjerrum length may be redefined as $q^2 \ell_{\mathrm{B}}$. 

\begin{figure*}[t]\begin{center}
	\begin{minipage}[b]{0.4\textwidth}\begin{center}
		\includegraphics[width=\textwidth]{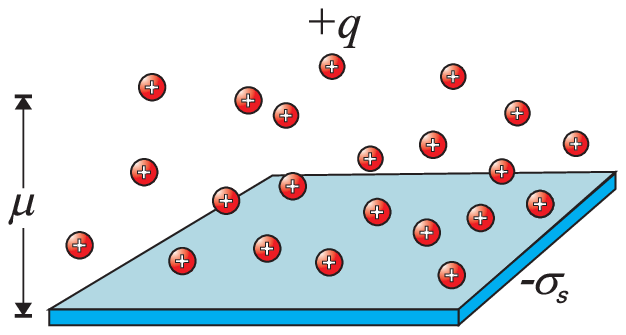} (a)
	\end{center}\end{minipage} \hskip0.35cm
	\begin{minipage}[b]{.4\textwidth}\begin{center}
		\includegraphics[width=\textwidth]{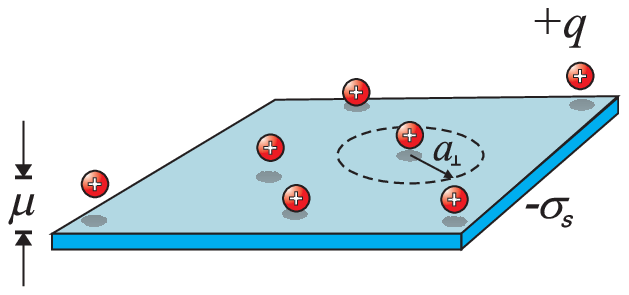} (b)
	\end{center}\end{minipage} \hskip0.35cm
	\caption{
	Schematic view of the structure of a layer of counterions at an oppositely
charged surface for small (a) and large (b) value of the coupling parameter. 
	 }
	\label{fig:onewall_layers}
\end{center}\end{figure*}

Other length scales may be identified by considering the specific form of the
charge distribution and geometry of macroions. 
For uniformly charged planar 
surfaces (Fig. \ref{fig:onewall_layers}), one can define another key length scale
by comparing the thermal energy with the energy scale of the counterion-wall attraction, i.e.,   $u/(k_{\mathrm{B}}T)=z/\mu$,  where  $z$ is the distance
from the wall and 
\begin{equation}
  \mu = {1}/({2\pi q \ell_{\mathrm{B}} \sigma_{\mathrm{s}}})
\label{eq:mu_wall}
\end{equation}
is known as the {\em Gouy-Chapman (GC) length} \cite{Gouy,Chapman}. 
The GC length measures the distance at which the thermal 
energy equals the counterion-wall electrostatic interaction energy. It also  gives a measure of the 
{\em thickness} of the counterion layer at a charged wall as we shall see later.

For planar systems where no other length scales are present, it follows that 
only the dimensionless ratio between the above two length scales matters, i.e., 
\begin{equation}
  \Xi={q^2 \ell_{\mathrm{B}}}/{\mu}=2\pi q^3 \ell^2_{\mathrm{B}} \sigma_{\mathrm{s}}. 
\label{eq:Xi}
\end{equation}
This parameter is known as the {\em electrostatic coupling parameter}  \cite{Netz01,AndreNetz}.

%%%%%%%%%%%%%%%%%%%%%%%%%%%%%%%%%%%%%%%%%%%%%%%%%%%%%%%%%%%%%%%%%%%%
 \section{From mean-field to strong coupling regime}
\label{subsec:MF_to_SC}

For small coupling parameters $\Xi  \ll 1$, equation (\ref{eq:Xi}) shows that 
the GC length is  relatively large, which indicates  that counterions  
form a loosely bound cloud at an oppositely charged wall 
(Fig. \ref{fig:onewall_layers}a). For large coupling parameter $\Xi\gg 1$, in contrast, the GC length is relatively small 
and counterions are strongly attracted toward the wall (Fig. \ref{fig:onewall_layers}b). Further insight 
may be obtained  by considering the typical distance between counterions 
at a charged surface. For counterions residing near the surface, the local electroneutrality 
condition implies a typical lateral separation of 
\begin{equation}
   a_\bot\sim \sqrt{{q}/{\sigma_{\mathrm{s}}}},
\label{eq:abot}
\end{equation}
since each  counterion neutralizes the charge of an area that scales as 
$a_\bot^2\sim q/\sigma_{\mathrm{s}}$. Counterion spacing $a_\bot$ is
not an independent length scale and may be written in the units of the GC length as
\begin{equation}
  ({a_\bot}/{\mu})\sim \sqrt{\Xi}. 
 \label{eq:abot_mu_Xi}
\end{equation}

\subsection{Weak coupling or mean-field regime}
\label{subsec:onewall_PB}

In the regime where $\Xi\ll 1$, equation (\ref{eq:abot_mu_Xi}) shows that the 
lateral separation of counterions near the surface is small compared with the
typical layer thickness, $\mu$, which  further indicates that the counterions tend to form a diffuse fluid-like layer at the surface (Fig. \ref{fig:onewall_layers}a).\footnote{A more accurate estimate 
of the typical distance, $a$,  between counterions  
in an extended three-dimensional layer gives $a/\mu \sim \Xi^{1/3}$ \cite{Netz01}.}  
This regime is dominated by collective mean-field-like effects, i.e.,  
counterions become uncorrelated from each other in a statistical sense 
as each counterion in the layer  interacts with a diffuse cloud of many other counterions. Therefore, $\Xi\ll 1$ identifies the {\em weak coupling} (WC) or 
{\em mean-field regime}, which is relevant to systems with weakly 
charged surfaces, low valency counterions and/or high temperature  \cite{hoda_review}. 

 Formally,  one can employ a mean-field approximation in order to describe
the system in the WC regime by neglecting all inter-particle correlations on the leading order. The mean-field approximation is exact in  the  limit $\Xi\rightarrow 0$ \cite{Netz-orland} and 
leads to the so-called Poisson-Boltzmann (PB) equation for the mean-field electrostatic potential 
$\psi({\mathbf r})$ \cite{VO,Israelachvili}, i.e.
\begin{equation}
  -\varepsilon \varepsilon_0\nabla^2 \psi= \rho_0({\mathbf r})
         +q e_0 n_0 \, \Omega({\mathbf r}) \,\exp(-\beta q e_0  \psi),
\label{eq:PBeq}
\end{equation}
where  $\beta = 1/ (k_{\mathrm{B}}T)$. The first term on the r.h.s.  represents the charge distribution due to fixed external charges (macroions),
 $\rho_0({\mathbf r})$, and
the second term is the mean-field number density of counterions 
$  n_{\mathrm{PB}}({\mathbf r})=n_0\,\Omega({\mathbf r})\, \exp(-\beta q e_0 \psi), $
where $n_0$ is a normalization prefactor. The  ``blip" function $\Omega({\mathbf r})$ is equal to one in the region accessible to counterions and zero elsewhere.

For  point-like counterions at a single uniformly charged wall, 
the PB equation yields the well-known algebraically decaying density profile  \cite{VO,Israelachvili}
\begin{equation}
  \frac{n_{\mathrm{PB}}(z)}{2\pi\ell_{\mathrm{B}}\sigma_{\mathrm{s}}^2}=\frac{1}{(z/\mu+1)^2}\quad \qquad z>0, 
\label{eq:onewall_PBdens}
\end{equation}
where $z$ is the distance from the wall. Note that the density profile is normalizable to the total number of counterions in order to ensure electroneutrality, and that the GC length, $\mu$, equals 
the height of a  layer containing {\em half} of the counterions,  thus giving a measure of the typical layer thickness. The {\em contact} density of counterions 
$n_{\mathrm{PB}}(z=0)= 2\pi\ell_{\mathrm{B}}\sigma_{\mathrm{s}}^2$ turns out to be an exact result within 
the present  model and remains valid beyond the mean-field theory \cite{contact_value}.

\subsection{Strong coupling regime}
\label{subsec:onewall_SC}

In the {\em strong coupling (SC) regime} $\Xi\gg 1$, equation (\ref{eq:abot_mu_Xi})  shows that the lateral separation of counterions becomes larger than the GC length, thus indicating that the counterions tend to form a quasi-2D layer at the surface (Fig. \ref{fig:onewall_layers}b). Such a layer is dominated by strong mutual repulsions between counterions as can be seen by considering the effective 2D plasma parameter \cite{Baus}
$  \Gamma \equiv {q^2 \ell_{\mathrm{B}}}/{a_\bot}\sim \Xi^{1/2}$, which  gives the ratio between Coulombic inter-particle repulsions and the thermal energy.
For elevated $\Xi$, Coulombic repulsions tend to freeze out  lateral fluctuations of counterions on the surface, leading  
to {\em strong correlations}  and  a trend toward crystallization in the ionic structure 
\cite{Rouzina96,Shklovs02}. Individual counterions thus become  isolated  in 
relatively large {\em correlation holes} of size $a_\bot$  from which neighboring counterions are statistically depleted. The Wigner crystallization of the 2D one-component plasma is  known to occur for
$\Gamma>\Gamma_c\simeq 125$ \cite{Baus}, which corresponds to the range of coupling parameters
$\Xi>\Xi_c\simeq 3.1\times 10^4$ \cite{AndreNetz}.

%-------------------------------------------------------------------------
%  Table 1
%-------------------------------------------------------------------------
%\begin{sidewaystable*}
\begin{table}[t]
\begin{center}
%Illustrative examples of physical parameters for some typical systems. 
\begin{tabular*}{10cm}{l  | c | c |  cl }
\hline\hline
 & $\sigma_{\mathrm{s}}$ ($e_0/$nm$^2$) & $q$ & $\mu$(\AA) &\qquad  $\Xi$  
\\ \hline
 charged membranes & $\sim 1$ & 1 & 2.2 &\qquad 3.1  \\
&  & 2 & 1.1 & \qquad 24.8  \\
&  &  3 & 0.7 & \qquad 83.7   \\
\hline
 DNA  & 0.9  & 1  (Na$^+$) & 2.4 & \qquad 2.8   \\
&  & 2  (Mn$^{2+}$) & 1.2 & \qquad 22.4     \\
&   & 3  (spd$^{3+}$) & 0.8 & \qquad 75.6     \\
&   & 4  (sp$^{4+}$) & 0.6 & \qquad 179   \\
\hline
 highly charged colloids  & $\sim 1$ &  3  & 0.7 & \qquad 85    \\
 (surfactant micelles) &  & & & \qquad \\
\hline
weakly charged colloids  &  $\sim 0.1$ &1& $\sim 2$  \qquad &\qquad $\sim 0.1$      \\
 (polystyrene particles)  &  & && \qquad \\
\hline
\end{tabular*}
\caption{
\label{tab:real_parameters}
Typical values of physical parameters for some realistic systems. }
\end{center}
%\vspace*{5mm}
%\end{sidewaystable*}
\end{table}

%-------------------------------------------------------------------------

For $\Xi\gg 1$, the PB description completely breaks down, nonetheless, 
one can obtain a simple analytical theory by employing a virial 
and $1/\Xi$ expansion to the leading order, which is known as
the {\em strong coupling theory} \cite{Netz01}.  
The SC theory turns out to 
%possess a simple analytical structure because it 
contain contributions that involve only single-particle interaction energies between counterions and 
the fixed macroion surface charges. For instance,  the SC density 
profile of counterions at a single charged wall comes exclusively from the vertical degree of freedom, $z$, through which 
single isolated counterions are coupled to the wall with the interaction potential $u/(k_{\mathrm{B}}T)=z/\mu$.   Hence using the Boltzmann weight, one finds  the (laterally averaged)  density profile 
\begin{equation}
  n_{\mathrm{SC}}(z)=n_0\exp(-z/\mu),
\label{eq:onewall_SCdens}
\end{equation}
where the prefactor (contact density) is again found from the electroneutrality
condition to be   $n_0= 2\pi\ell_{\mathrm{B}}\sigma_{\mathrm{s}}^2$. Unlike in the WC case, 
the SC density profile  decays exponentially away from the charged wall. Moreover, the average distance of counterions  is finite and equal to the GC length, $\langle z\rangle_{\mathrm{SC}} =\mu$, reflecting again the quasi-2D structure of the layer. 

Formally, the single-particle SC theory is exact in the asymptotic limit of an infinitely large correlation hole size, $a_\bot/\mu\rightarrow \infty$,  or simply $\Xi\rightarrow \infty$. However, its validity holds in a wider range of system parameters as is evident from comparison with computer simulations \cite{Netz01,AndreNetz}. For instance, for a {\em finite} coupling parameter $\Xi$, the SC density profile (\ref{eq:onewall_SCdens}) still holds approximately  at  distances  $z<a_\bot$, which yields the criterion
\begin{equation}
  {z}/{\mu}<\sqrt{\Xi},
 \label{eq:crit_onewall}
\end{equation}
identifying the limits of applicability of the SC theory. At larger distances $z>a_\bot$, 
multi-particle interactions become increasingly more important and the
mean-field-like features eventually dominate for  $z\gg a_\bot$ \cite{Netz01,AndreNetz}.

In brief, thus, one can identify two asymptotic regimes of weak and strong  
 coupling based on the value of  the electrostatic coupling parameter,
 where a charged system may be studied by means of two limiting theories, 
 namely,  the WC (mean-field) and the SC theory. 

In Table \ref{tab:real_parameters}, we present illustrative examples of the 
parameter values (surface charge density $\sigma_{\mathrm{s}}$, 
counterion valency $q$, GC length $\mu$, and the coupling parameter $\Xi$ in
water and at room temperature) from a few realistic weakly and strongly  coupled  systems. 
Note that a typical coupling strength of $\Xi\sim 10^2$ (or larger) already falls within the SC regime 
and a value of $\Xi\sim 1$ (or smaller) typically well inside the WC 
regime \cite{Netz01,AndreNetz,hoda_review,Naji_PhysicaA}.

%%%%%%%%%%%%%%%%%%%%%%%%%%%%%%%%%%%%%%%%%%%%%%%%%%%%%%%%%%
 \section{Interactions between like-charged surfaces}
\label{sec:interaction}

Macroions in solution are often like-charged and thus repel each other by their bare Coulomb 
interaction. The overall interaction is however different from this bare interaction due
to the presence of counterions, which can mediate both {\em repulsive} and {\em attractive} effective
forces. Obviously, the counterion-mediated interactions depend strongly on the distribution of counterions
around macroions.

\begin{figure*}[t]\begin{center}
	\begin{minipage}[b]{0.43\textwidth}\begin{center}
		\includegraphics[width=\textwidth]{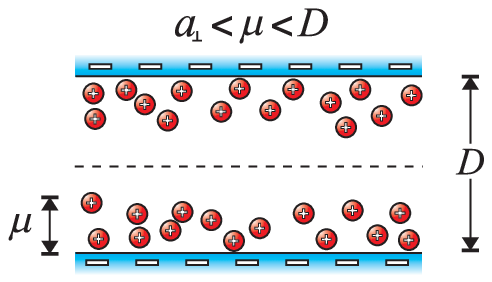} (a)
	\end{center}\end{minipage} \hskip0.35cm
	\begin{minipage}[b]{.37\textwidth}\begin{center}
		\includegraphics[width=\textwidth]{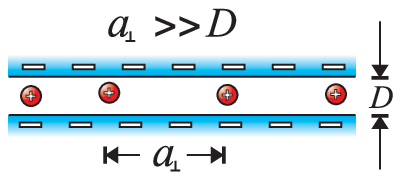} (b)
	\end{center}\end{minipage} \hskip0.35cm
	\caption{
Schematic representation of the WC (a) and  SC (b) interaction regimes for two like-charged walls. 
	 }
	\label{fig:twowalls_PBSC}
\end{center}\end{figure*}

In order to demonstrate the underlying physical picture, we shall focus on the interaction between two 
{\em identical} planar like-charged walls of uniform surface charge density $-\sigma_{\mathrm{s}} e_0$ at separation $D$, where $q$-valent counterions fill only the space between the walls, Fig. \ref{fig:twowalls_PBSC} (the dielectric constant is assumed to be uniform in space).  
In this system, an extra length scale is set by the wall separation, $D$. 
Two limiting regimes of  repulsion and attraction may be distinguished qualitatively 
by comparing $D$ with other length scales of the system as follows.

\subsection{WC regime: Repulsion}
\label{subsec:twowalls_PB}

Let us first consider the WC limit, $\Xi\ll 1$, in the regime where the wall separation, $D$, is large
compared with all other length scales (Fig. \ref{fig:twowalls_PBSC}a).
In this case, counterions form a diffuse layer at each wall,
but due to the large separation, the system is approximately decoupled into two 
nearly neutral sub-systems, each consisting of a charged wall and its 
counterion cloud. The effective pressure acting between the walls is dominated 
by the osmotic pressure of counterions across the mid-plane, which is positive
and thus corresponds to an effective {\em repulsion} between the walls. 
The scaling behavior of this effective repulsion follows by noting 
that the mid-plane osmotic pressure can be estimated from 
the local counterion density,
$n_{\mathrm{mid}}$, and by using the ideal-gas equation of 
state as $\beta P\sim n_{\mathrm{mid}}$. Thus, according 
to  Eq. (\ref{eq:onewall_PBdens}),  the interaction pressure is expected to decay
as $\sim D^{-2}$. 

It turns out that the PB pressure obtained in the limit $\Xi\rightarrow 0$ coincides {\em exactly} with the mid-plane osmotic pressure of counterions as 
discussed above and may be expressed  as  
 \cite{VO,Israelachvili,Netz01}
\begin{equation}
    \frac{ \beta P_{\mathrm{PB}}(D)}{2\pi \ell_{\mathrm{B}} \sigma_{\mathrm{s}}^2} =\Lambda, 
\label{eq:P_PBfull}
\end{equation} 
where $\Lambda$ is obtained from $\Lambda^{1/2} \tan[\Lambda^{1/2} (D/2\mu)] = 1$. From here one can obtain the large-separation $D/\mu\gg 1$ behavior 
\begin{equation}
  \frac{ \beta P_{\mathrm{PB}}(D)}{2\pi \ell_{\mathrm{B}} \sigma_{\mathrm{s}}^2}
          \simeq \left(\frac{\pi\mu}{D}\right)^2. 
\label{eq:P_PB}
\end{equation}

\subsection{SC regime: Attraction}
\label{subsec:twowalls_SC}

Now let us consider the SC limit, $\Xi\gg 1$, in the regime where $D$ is  smaller than the lateral spacing between counterion,  $a_\bot\gg D$ (Fig. \ref{fig:twowalls_PBSC}b). Since counterions are highly separated from each other, the two opposite layers of counterions tend to form an inter-locking pattern 
at small separations.  

The system may be thought of as a collection of laterally frozen ``correlation cells'', each
 consisting of a single counterion sandwiched between 
two opposing sections of the walls with lateral size of about $a_\bot$. Since $a_\bot\gg D$, 
the effective pressure  between the walls is dominated 
by the contribution from single counterions  fluctuating in single correlation cells.  
The electrostatic energy of the system per cell is the sum of the 
bare interactions between the two surfaces with each other 
and with the single counterion, which--using the electroneutrality 
condition per cell and the fact that the wall separation is small--follows as 
$ u_{\mathrm{elec}}/(k_{\mathrm{B}}T)\simeq 2\pi\ell_{\mathrm{B}}\sigma_{\mathrm{s}}^2 D$ per unit area. 
This {\em energetic} contribution gives an attractive pressure as 
$ \beta P_{\mathrm{elec}}\simeq -2\pi\ell_{\mathrm{B}}\sigma_{\mathrm{s}}^2$ between the walls. 
On the other hand, the entropic contribution due to counterion confinement 
is of the order $S_{\mathrm{ci}}\sim k_{\mathrm{B}}\ln D$ (per cell), which generates 
a repulsive component. The total pressure between strongly coupled walls is then obtained by combining these two effects as 
\begin{equation}
  \frac{ \beta P_{\mathrm{SC}}(D)}{2\pi\ell_{\mathrm{B}}\sigma_{\mathrm{s}}^2}=
      -1+\frac{2\mu}{D}.
\label{eq:P_SC}
\end{equation}
This expression clearly predicts a closely packed {\em bound state} for the like-charged walls
with an equilibrium surface-surface separation, $D^\ast$, equal to twice the GC length, i.e., 
$  D^\ast=2\mu$. The like-charged walls therefore {\em attract} each other for $D>D^\ast$ and {\em repel} at smaller separations.

The analytical expression  (\ref{eq:P_SC}) 
is indeed an exact  result for planar walls in the limit $\Xi\rightarrow \infty$ \cite{Netz01,AndreNetz}.  It turns out that in a system with {\em finite} coupling parameter, $\Xi$, 
the asymptotic ($\Xi\rightarrow \infty$)  SC  results 
still hold approximately as long as the surface separation, $D$, is smaller than the typical lateral 
distance between counterions, $a_\bot$, i.e.,  for
$  D<a_\bot$. This condition in fact yields a simple and generic criterion identifying the regime where the SC attraction is expected to emerge between two like-charged macroions. 
It was originally suggested by Rouzina and Bloomfield \cite{Rouzina96} and verified and generalized later 
using extensive analytical and numerical methods \cite{Netz01,AndreNetz,hoda_review,Naji_PhysicaA,jho-prl,Naji_CCT, SCdressed,Matej-cyl,Naji-cylinders,asim}. 
For larger inter-surface separations, $D>a_\bot$, the mean-field features
 become increasingly more important and the strength of attraction reduces. Eventually at very large
$D$, the interaction becomes repulsive \cite{Netz01,AndreNetz}.

%%%%%%%%%%%%%%%%%%%%%%%%%%%%%%%%%%%%%%%%%%%%%%%%%%%%%%%%%%%%%%%%%%%%
\section{Counterions with salt}

The SC theory was so far designed exclusively for counterions-only systems,  i.e., Coulomb fluids composed of only counterions in the absence of any salt ions \cite{Netz01}. Though an approximation of this type can be used to describe situations where a large amount of polyvalent counterions dominate the system, it has to be amended in the general case in order to deal with the complexity of real systems that always contain some amounts of simple salt  \cite{Israelachvili}. An experimentally oft-encountered situation would be a system composed of fixed surface charges with polyvalent counterions bathed in a solution of univalent salt \cite{rau-1,rau-2}. 

This situation leads to a difficult problem of {\em asymmetric} aqueous electrolytes where different components of the Coulomb fluid are differently coupled to local electrostatic fields \cite{olli}. Polyvalent counterions are coupled strongly, whereas univalent salt ions are coupled weakly. In this case no single approximation scheme that would treat all the charged components on the same level would be expected to work. Whereas the SC framework would certainly work for the polyvalent counterions, it would fail for the univalent salt. The converse is true for the WC framework. One is thus faced with a problem since no single approximation scheme appears to be valid in any range of coupling parameters. One can nevertheless build a theoretical framework that allows to {\em selectively} use different approximation schemes for different components of the asymmetric Coulomb fluid. This combined WC-SC approach appears to bring forth all the salient features of these asymmetric systems at high electrostatic couplings \cite{SCdressed}. 

\subsection{Functional integral formalism}

Our arguments until now were strictly intuitive. A formal theory can be developed exactly in terms of the functional integral representation of the classical partition function of the Coulomb fluid along the lines first introduced by Edwards and Lenard \cite{Edwards,podgornik,Netz01}. 

Assume first that the system is composed of charged macromolecules with fixed charge density $\rho_0(\vct r)$, mobile polyvalent counterions and an additional univalent salt. The total electrostatic interaction energy of  a given configuration of the system can be written as
\begin{equation}
W=\frac 12 \int\rho(\vct r)v(\vct r, \vct r')\rho(\vct r')\,\rmd \vct r\,\rmd\vct r',
\label{energy_new_1}
\end{equation}
where $v(\vct r, \vct r')$ is the Coulomb kernel given by 
$v(\vct r, \vct r')={1}/({4\pi\varepsilon\varepsilon_0\vert\vct r-\vct r'\vert})$, and $\rho(\vct r)$ is  the total charge density 
\begin{equation}
\rho(\vct r)=\rho_0(\vct r)+\sum_i q e_0\delta(\vct R_i^c\!\!-\vct r)+\sum_i e_0\delta(\vct R_i^+\!\!-\vct r)-\sum_i e_0\delta(\vct R_i^-\!\!-\vct r), 
\end{equation}
where $\vct R_i^{c}$, $\vct R_i^{+}$ and $\vct R_i^{-}$ are the positions of the polyvalent counterions, univalent cations (salt counterions) and univalent anions (salt coions), respectively.\footnote{We may refer to the $q$-valency (polyvalent)  counterions simply as  ``counterions".}  
Assuming again that the system is composed of two apposed planar surfaces  at $z = \pm D/2$  with the surface charge density $-\sigma_{\mathrm{s}}$, we have 
\begin{equation}
\rho_{0}(\vct r) = -\sigma_{\mathrm{s}}e_0 \big[\delta(z - D/2) + \delta(z + D/2) \big].  
\label{eq:rho_0} 
\end{equation}
The salt ions are assumed to be present in all regions in space, whereas the counterions are assumed
to be present only in the inter-surface region $|z|<D/2$ as specified by the geometry ``blip" function $\Omega (\vct r)$ (Section (\ref{subsec:onewall_PB})).

We then follow the standard procedure by introducing a fluctuating local potential, $\phi$, via the Hubbard-Stratonovich transformation, which leads to the following exact functional integral representation for the grand-canonical partition function \cite{Netz01,podgornik}
\begin{equation}
{\mathcal Z}=\int{\mathcal D}\phi\,\rme^{-\beta H[\phi]},
\label{partfun}
\end{equation}
where the field-functional Hamiltonian reads
\begin{eqnarray}
H[\phi]&=&\frac 12\int\phi(\vct r)v^{-1}(\vct r,\vct r')\phi(\vct r')\,\rmd\vct r\,\rmd\vct r'  + \rmi\int\rho_0(\vct r)\phi(\vct r)\, \rmd \vct r \label{hamiltonian}
\\
&-&\frac{\Lambda_c}{\beta}\int\rme^{- \rmi\beta q e_0\phi(\vct r)}\Omega(\vct r)\, \rmd \vct r  
- \frac{\Lambda_+}{\beta}\int\rme^{-\rmi\beta e_0\phi(\vct r)}\, \rmd \vct r
- \frac{\Lambda_-}{\beta}\int\rme^{\rmi\beta e_0\phi(\vct r)}\, \rmd \vct r, 
\nonumber
\end{eqnarray}
and $\Lambda_c$ and $\Lambda_{\pm}$ represent the fugacities of polyvalent counterions and salt ions and $v^{-1}(\vct r,\vct r')=-\varepsilon\varepsilon_0 \nabla^2 \delta(\vct r-\vct r')$ is the inverse Coulomb kernel. 
The special case of counterions-only system, as analyzed in the previous section, is recovered by setting $\Lambda_{\pm} = 0$. 

%Though the derivation of these results need not be strictly tied to the functional integral representation of the partition function
%it provides a convenient gateway to further generalization where it plays an essential role. 

We shall assume that salt ions are in equilibrium with a bulk reservoir
containing equal concentration $n_b$ of both positive and negative ions, which implies  $\Lambda_+=\Lambda_-\equiv n_b$. One can thus introduce  the Debye-H\" uckel (DH) screening parameter $\kappa$ (inverse ``screening length") as $\kappa^2=8\pi \ell_{\mathrm{B}} n_b.$
The polyvalent counterions shall be treated here within the canonical ensemble \footnote{see Ref. \cite{SCdressed} for a grand-canonical description of polyvalent counterions.}, i.e., their number in the slit is assumed to be fixed and equal to $N$. 
The number of counterions may be expressed via the dimensionless parameter 
\begin{equation}
\eta={Nq }/({2\sigma_{\mathrm{s}} S}),
\label{eta}
\end{equation}
where $S$ is the area of the interacting surfaces. The case $\eta=0$ represents a system with salt only, and $\eta=1$ is the case where the total charge due to counterions exactly compensates the surface charge. Note that  $\eta$ can take any non-negative value when salt ions are present. This is because salt ions turn the long-range Coulomb potential into a short-range DH potential (see below) and can thus ensure the electroneutrality condition themselves.

\subsection{Dressed counterions}

Assuming that the system is highly asymmetric $q \gg 1$, one can formulate an approximate theory in order to evaluate the partition function (\ref{partfun}) analytically by acknowledging the fact that 
the polyvalent counterions are strongly coupled while the simple salt ions are weakly coupled to the fluctuating electrostatic fields. This leads to a mixed WC-SC evaluation of the partition function \cite{SCdressed}. 

The salt terms (the last two terms) in Eq. (\ref{hamiltonian}) can be combined into
$\cos\,\beta e_0\phi(\vct r)$ and in a highly asymmetric system this can be expanded up to the quadratic 
order in the fluctuating potential. 
%is sensible and becomes exact in the limit of $q \gg 1$. 
Thus up to an irrelevant constant we remain with an effective field Hamiltonian of the form
\begin{eqnarray}
H_{\mathrm{eff}}[\phi]&=&\frac 12\int\phi(\vct r)v_{\rm DH}^{-1}(\vct r,\vct r')\phi(\vct r')\,\rmd\vct r\,\rmd\vct r' + \rmi\int\rho_0(\vct r)\phi(\vct r)\,\rmd \vct r\nonumber\\
&-&\frac{\Lambda_c}{\beta}\int\rme^{- \rmi\beta q e_0\phi(\vct r)} \Omega(\vct r)\,\rmd \vct r.
\label{hamiltonian-renorm}
\end{eqnarray}
This procedure therefore yields an effective Hamiltonian for a ``counterions-only" system but with the proviso that the inverse Coulomb kernel is replaced by the standard inverse DH kernel 
\begin{equation}
v_{\rm DH}^{-1}(\vct r,\vct r')=-\varepsilon\varepsilon_0(\nabla^2-\kappa^2)\delta(\vct r-\vct r'), \qquad {\rm with} \qquad v_{\rm DH}(\vct r,\vct r')=\frac{\rme^{-\kappa \vert\vct r-\vct r'\vert}}{4\pi\varepsilon\varepsilon_0\vert\vct r-\vct r'\vert}.
\label{DHinv}
\end{equation}
We have thus effectively integrated out the salt degrees of freedom leading to a renormalized interaction potential between all the remaining charge species of the screened DH form. One can thus drop any reference to explicit salt ions and infer the thermodynamic properties of the original system by analyzing it as a system composed of   {\em dressed counterions} and fixed external charges interacting via a screened DH pair potential. 
In the SC limit for the polyvalent counterions we term this approximation scheme as the {\em SC dressed counterion theory}. Our SC analysis thus proceeds in the same way as for the 
counterions-only systems \cite{Netz01} except that the interactions
between the charges are now of a dressed form. 

We note that any Bjerrum pairing \cite{Bjerrum,vanroij} or even electrostatic collapse of the salt or formation of salt-counterion complexes \cite{Fisher} is beyond the framework developed here.

\subsection{WC dressed counterion theory}
\label{subsec:WCdressed}

We again focus on a system composed of two plane-parallel surfaces defined via
Eq. (\ref{eq:rho_0}). In the WC limit (for both the counterions as well as the salt ions), 
the functional integral derived in the previous section is dominated by the contribution
from the saddle-point solution  $\phi_{\mathrm{SP}}$. 
%\begin{equation}
%\delta H[\phi]/\delta \phi|_{\phi_{\mathrm{SP}}}=0.
%\end{equation}  
This subsequently leads to the mean-field equation for the real-valued mean-field potential $\psi=\rmi \phi_{\mathrm{SP}}$, i.e., 
\begin{equation}
 - \varepsilon\varepsilon_0 (\nabla^2\psi - \kappa^2\psi)= \rho_0({\mathbf r}) + q e_0 \Lambda_c  \Omega(\vct r) \,\rme^{-\beta q e_0 \psi}, 
 \label{PB_0}
\end{equation}
which, in rescaled units $w= \beta q e_0 \psi$ and by virtue of 
the lateral symmetry  for planar surfaces $w=w(z)$, may be written as
\begin{equation}
w''=\kappa^2 w-C\, \rme^{-w} \quad \qquad \vert z\vert<D/2. 
\label{canon-PB}
\end{equation}
The constant $C$ can be evaluated when one stipulates the fixed amount of counterions. Outside the slit $\vert z\vert>D/2$, the mean-field equation has the standard DH form $w''=\kappa^2 w$, which yields $w(z)=w_0\exp(\pm \kappa z)$.

The interaction pressure, $P$, between the bounding surfaces is given by the difference of the ion concentrations at the mid-plane ($z=0$), where the mean electric field vanishes, and the bulk concentration, i.e., $\beta P=n_+(0)+n_-(0)+n_c(0)-2n_b$, which leads to the dimensionless expression 
\begin{equation}
\frac{ \beta P(D)}{2\pi\ell_{\mathrm{B}}\sigma_{\mathrm{s}}^2} = \frac 14 (\kappa \mu)^2\, w^2(0)+\frac{1}{2}C\, \rme^{-w(0)}. 
\label{eq:PB_DH_C}
\end{equation}
As evident from the above equation, the pressure can never be negative and the effective interaction 
is thus {\em always repulsive} within this type of mean-field approach \cite{PBrepulsive}. 
The canonical mean-field equation  (\ref{canon-PB}) can be solved numerically and the results can be used to evaluate the pressure (\ref{eq:PB_DH_C}) as a function of the inter-surface separation (Fig. \ref{fig_press2}a, dashed line). 

\subsection{SC dressed counterion theory}

The analysis of the dressed counterions system in the SC limit is very similar to the traditional SC approach in 
the case of counterions only \cite{Netz01}. We proceed by 
expanding the grand-canonical partition function associated with the dressed counterion approximation, Eq. (\ref{hamiltonian-renorm}), to the first order in counterion fugacity, $\Lambda_c$.  We then perform an inverse Legendre transformation \footnote{This is achieved by mapping the fugacity to the
number of counterions, $N$, via the relation $\Lambda_c\partial \ln {\mathcal Z}/\partial \Lambda_c = N$.} in order to obtain the canonical SC free energy \cite{Naji-cylinders,Netz01}. 
We thus find
\begin{equation}
{\mathcal F}_{N}= U_0-Nk_{\mathrm{B}}T\,\ln\int \rme^{-\beta u(\vct r)}\,\rmd \vct r,
\label{SCfree}
\end{equation}
where the first term is the screened interaction energy of fixed charges 
\begin{equation}
U_0=\cfrac{1}{2}\int \rho_0(\vct r)v_{\mathrm{DH}}(\vct r,\vct r')\rho_0(\vct r')\,\rmd \vct r\,\rmd \vct r',
\label{eq:W_00}
\end{equation}
and the term in the exponent is 
the single-particle interaction energy of the dressed counterions with fixed macroion charges
\begin{equation}
u(\vct r)= q e_0 \!\int  \rho_0(\vct r')v_{\mathrm{DH}}(\vct r,\vct r')\,\rmd \vct r'.
\label{W0c}
\end{equation}
The SC attraction between like-charged macroions stems from the second term in Eq. (\ref{SCfree}), which contains the counterion-induced effects \cite{Naji_PhysicaA,hoda_review,Naji-cylinders,Netz01}. 

For the planar system under consideration (Eq. (\ref{eq:rho_0})), the above quantities may be evaluated explicitly \cite{SCdressed}
and we find  the SC free energy  as 
\begin{equation}
\beta {\mathcal F}_{N}/N=\frac{1}{2\kappa\mu\eta}\,\rme^{-\kappa D}-\ln\,I(D),
\label{FSC}
\end{equation}
where we have introduced
\begin{equation}
I(D)=\int_0^{D/2}\exp\Bigl(\frac{2}{\kappa\mu}\,\rme^{-\kappa D/2}\cosh\,\kappa z\Bigr)\,\rmd z.
\label{I}
\end{equation}
The first term in Eq. (\ref{FSC}) corresponds to the usual salt-mediated  repulsive DH interaction between the two surfaces, and the second one is the  contribution of counterions, which is 
proportional to $\eta$, Eq. (\ref{eta}), on the single-particle SC level. 

The dimensionless pressure acting between the surfaces  can be obtained from the free energy via the standard thermodynamic relation $P=-\partial({\mathcal F}_{N}/S)/\partial D$, thus yielding 
\begin{equation}
 \frac{ \beta P(D)}{2\pi\ell_{\mathrm{B}}\sigma_{\mathrm{s}}^2} =\rme^{-\kappa D}+2\eta\,\frac{I'(D)}{I(D)},
\label{p_C}
\end{equation}
where the prime denotes the derivative with respect to the argument. 

\begin{figure*}[t]\begin{center}
	\begin{minipage}[b]{0.43\textwidth}\begin{center}
		\includegraphics[width=\textwidth]{press04-5new.eps} (a)
	\end{center}\end{minipage} \hskip0.25cm
	\begin{minipage}[b]{0.51\textwidth}\begin{center}
		\includegraphics[width=\textwidth]{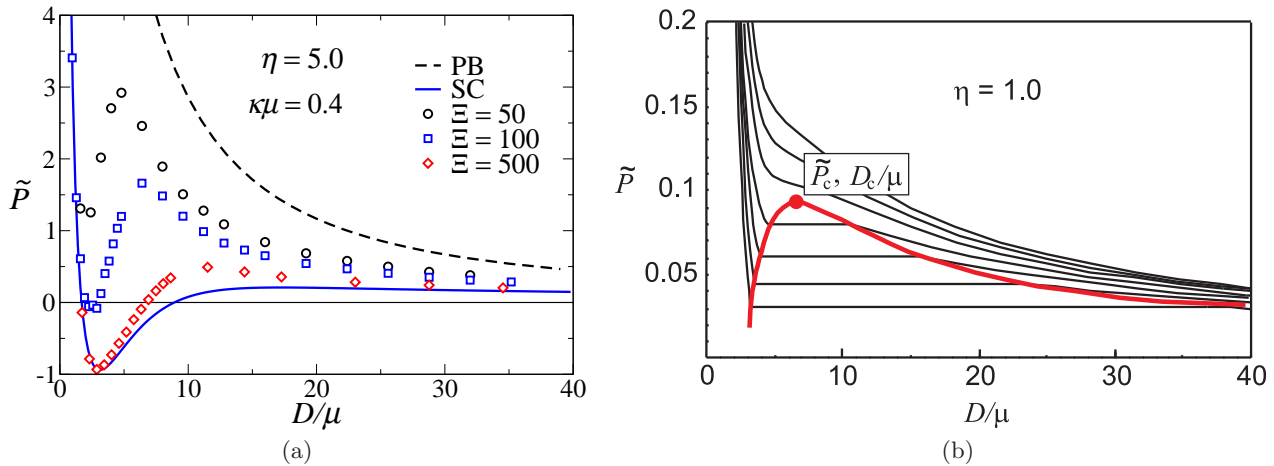} (b)
	\end{center}\end{minipage} \hskip0.25cm
	\caption{
	a) Rescaled interaction pressure $\tilde P \equiv  \beta P(D)/(2\pi\ell_{\mathrm{B}}\sigma_{\mathrm{s}}^2)$ between two like-charged surfaces as a function of the rescaled inter-surface distance. 
	%Solid line shows the SC result, Eq. (\ref{p_C})), dashed line shows  the PB result,  Eq. (\ref{eq:PB_DH_C}), and the symbols are Monte-Carlo simulation data.  
	b) Van der Waals type iso-inverse screening length curves are shown along with the corresponding Maxwell construction. The inverse screening length is varied in the range from $\kappa\mu = 0.4$ to $\kappa\mu = 0.56$ in intervals of 0.027
 (from bottom to top). 
 %Red line represents the critical curve. 
 }
	\label{fig_press2}
\end{center}\end{figure*}

The analytical SC pressure (\ref{p_C})  is shown in  Fig.~\ref{fig_press2}a as a function of the inter-surface distance (solid curve) 
along with the results from Monte-Carlo simulations\cite{SCdressed} (symbols) for a few different screening parameters. We also show in the figure the WC pressure from Eq. (\ref{eq:PB_DH_C}) (dashed curve). First note that 
the simulation results are bracketed by the two limiting analytical theories of WC and SC within the dressed counterion scheme and thus agree with the general feature obtained before \cite{Naji_PhysicaA,Netz01,AndreNetz,hoda_review} that the WC and SC limits in fact establish the upper and lower bounds for the interaction pressure between charged surfaces.

For both small and large separations the interaction pressure becomes positive
(repulsive), whereas for a sufficiently large coupling parameter \cite{SCdressed}, an effective inter-surface attraction can emerge at intermediate separations between the walls.  For $\kappa D\ll 1$,  the SC pressure reduces to
\begin{equation}
\frac{ \beta P(D)}{2\pi\ell_{\mathrm{B}}\sigma_{\mathrm{s}}^2} \simeq\frac{2 \mu\eta}{D}+(1-2\eta), 
\label{psmall}
\end{equation}
which to the leading  order corresponds to the ideal-gas osmotic pressure of counterions squeezed between the two surfaces. For $\kappa D\gg 1$, the pressure behaves as
\begin{equation}
\frac{ \beta P(D)}{2\pi\ell_{\mathrm{B}}\sigma_{\mathrm{s}}^2}  \simeq\frac{2 \mu\eta}{D}-\frac{4\eta}{(\kappa D)^2},
\label{plarge}
\end{equation}
which indicates that at large separations the counterions again behave as an ideal gas on the leading order as all electrostatic interactions are effectively  screened out and hence only the repulsive osmotic contribution remains.

%The canonical dressed counterion theory can be used to derive the counterions-only SC pressure \cite{Netz01}, i.e., $\tilde P=1/\tilde a-1$  in Eq. (\ref{eq:P_SC}) by letting $\kappa\to 0$ and setting $\eta=1$ in order to satisfy the electroneutrality condition. 
% since for vanishing salt the electrostatic interactions are long ranged due to a lack of screening. 
%The corresponding pressure for the counterions-only case is $\tilde p=1/\tilde a-1$, coinciding exactly with the form derived by Netz \cite{Netz01}.

The interaction pressure in the canonical ensemble thus always possesses repulsive branches at small and large separations and can show non-monotonic behavior in between.  In fact, for certain values of the parameters the interaction pressure 
shows a van der Waals-like loop which could suggest a coexistence regime between two different   ``phases". This loop is obtained for certain iso-ionic strength curves. From a thermodynamic perspective one thus has a coexistence between a dense phase, identified with a small inter-surface separation, at equilibrium with an expanded phase with a larger inter-surface separation.  
Such van der Waals-like coexistence between interacting charged surfaces has been seen in other contexts before \cite{Orr} and can be demonstrated by means of a Maxwell construction analysis as shown in Fig. \ref{fig_press2}b. 

In an experiment such as osmotic stress experiments and surface force experiments \cite{Israelachvili} one can only probe stable equilibrium states of the system implying interaction pressure vs. separation curves that are in agreement with the appropriate Maxwell construction. The binodal or the coexistence curve, which delimits the region in the pressure-separation plots where a Maxwell construction is feasible  (red curve  in Fig. \ref{fig_press2}b), ends at a critical point corresponding to a critical amount of salt
above which the interaction pressure remains purely repulsive. For the case with $\eta=1$, we find the critical point as ($\tilde P_c=0.092$, $\ D_c/\mu=6.14$, $\kappa_c\mu=0.546$).

It is interesting to note that this type of interaction pressure equilibria corresponding to abrupt transitions from one equilibrium separation to a different one have been observed in experiments with strongly charged macromolecules in the presence of polyvalent counterions and univalent salt. A typical example would be the osmotic stress experiments on DNA in the presence of trivalent CoHex counterions and 0.25M NaCl salt, that show abrupt transitions in osmotic pressure for intermediate ${\rm CoHex}^{3+}$ concentrations  from one repulsive osmotic pressure branch to another one \cite{rau-1}. Similar features are discerned even for a divalent counterion  ${\rm Mn}^{2+}$ at various concentrations or temperatures \cite{rau-2}.

The agreement between the SC dressed counterion theory and simulations becomes better as the coupling parameter $\Xi$ becomes larger. The agreement is also better for a smaller fraction of counterions, $\eta$, in the slit. 
Using a similar argument as in the counterions-only case in Section \ref{subsec:twowalls_SC} \cite{hoda_review,Naji_PhysicaA,Netz01,AndreNetz}, we find that the theory is expected to hold at small separations given by
$D/\mu\ll \sqrt{\Xi/\eta}$. Thus, for $\eta<1$ (i.e., when
the amount of the bare charge due counterions is less than the bare fixed charge on the macroions), the SC dressed counterion theory is expected to hold in a wider range of separations as compared with the original counterions-only SC theory \cite{Netz01}. At very large separations, where most of the electrostatics is screened out, the interaction between counterions becomes negligible  and the SC theory of dressed counterions retains its validity again. This result is a consequence of the dressed counterion theory and is not obtained in the standard SC theory with counterions only. Thus, the SC dressed counterion theory captures the physics both at large and small separations but would require improvements at intermediate separations.  

It should be noted that the validity of the DH-type linearization  that we have used to derive the dressed interaction potentials is also limited by stipulating that the dimensionless DH potential itself is always small enough. This leads   to the condition that $\kappa\gg 2\pi\ell_{\mathrm{B}}\sigma_{\mathrm{s}}$, or 
$\kappa\mu\gg 1/q$, which turns out to cover a whole range of realistic parameter values \cite{SCdressed}.

%%%%%%%%%%%%%%%%%%%%%%%%%%%%%%%%%%%%%%%%%%%%%%%%%%%%%%%%%%%%%%%%%%%%
\section{Counterions between randomly charged surfaces}

The assumption of homogeneity of surface charges is in general quite severe and there are well known cases where this assumption is not realistic at all.
Random polyelectrolytes and polyampholytes present one such case
\cite{andelman-disorder,kantor-disorder1,kantor-disorder2}. There the sequence of charges can be
distributed along the polymer backbone in a disordered manner where the disorder distribution may be of a {\em quenched} type. The Coulomb (self-)interactions of such polyelectrolytes are distinct and different from homogeneously charged polymers.

A case even closer to the present line of reasoning are investigations of interactions between solid surfaces in the presence of charged surfactants. The aggregation of  surfactants on crystalline
hydrophobic substrates in aqueous solutions can sometimes show structures consistent with highly inhomogeneous and disordered surface charge distributions \cite{manne1}.
%with half-cylinders on  for quaternary ammonium surfactants (above the critical
%micelle concentration), full cylinders on mica, and spheres on amorphous silica  
%Such interfacial aggregates whose emergence and
%structural details depend on the method of preparation result
%from a compromise between the natural free curvature as defined by
%intermolecular interactions and the constraints imposed by specific
%surfactant-surface interactions and can  pattern interacting surfaces at
%nanometer-length scales. 
Similar interfacial structures are seen for interacting
hydrophilic mica surfaces in the presence of cetyl-trimethyl-amonium bromide
(CTAB) or other surfactant-coated surfaces \cite{klein}. The surfaces appear to be covered by a
mosaic of positively and negatively charged regions and experience a strong,
{\em long-ranged attraction}, which is comparable in magnitude to that between
hydrophobic surfaces, and  is orders of magnitude larger than the expected 
Lifshitz-van der Waals forces \cite{klein}.  The patterning of interacting surfaces described above is
highly disordered, depends on the method of preparation and has basic
implications also for the forces that act between other types of hydrophilic
surfaces with mixed charges. 

It thus seem appropriate to investigate the effect of
quenched disordered charge distribution on the interactions
between macroions in ionic solutions.\footnote{See Refs. \cite{partial,disorder-PRL} for an analysis of 
the effects due to annealed and partially annealed charge disorder.} 
As a particular case, we shall again focus on the effective interaction between two randomly charged planar surfaces   across a  one-component Coulomb fluid \cite{ali-rudi}.

\subsection{General formalism: The replica method}

The  partition function of a Coulomb fluid in the presence of an external fixed charge distribution  $\rho_0(\vct
r)$, can be again written in the form of a functional integral over the fluctuating
electrostatic field $\phi(\vct r)$ as given in Eq.  (\ref{partfun}). 
However, the fixed charges are now assumed to be randomly distributed 
on macromolecular surfaces. Thus $\rho_0(\vct r)$ is represented by a 
probability distribution, which is assumed to be Gaussian  with no spatial correlations, i.e.
\begin{equation}
{\mathcal P}[ \rho_0(\vct r)]  = {\mathrm{const.}} \times \rme^{- \ul12 \int
{\mathrm{d}}\vct r~g^{-1}(\vct r) \left( \rho_0(\vct r) - \bar \rho_0(\vct r)\right)^{2}},
\label{distr-1}
\end{equation}
where $\bar \rho_0(\vct r)$ is the mean value and $g(\vct r) $ the width or variance of the charge disorder distribution.
For clarity, we shall also focus on the counterions-only case by  setting $\Lambda_\pm=0$ in Eq. (\ref{hamiltonian}). 

The average over quenched charge disorder is now obtained by applying the standard
Edwards-Anderson replica ansatz \cite{orland,dotsenko} in the form
\begin{equation}
{\mathcal F} = - k_{\mathrm{B}} T \, \overline{\ln{{\mathcal Z}}} = - k_{\mathrm{B}} T\,\lim_{n \rightarrow 0}
\frac{\overline{{\mathcal Z}^{n}} - 1}{n}, 
\label{free-1}
\end{equation}
where the disorder average is defined as $\overline{(\dots)} = \int {\mathcal D}\rho_0\, (\cdots) \, {\mathcal P}[ \rho_0(\vct r)] $.

The Gaussian integrals involved in Eq. (\ref{free-1}) can be evaluated straightforwardly  and the final form of the replicated 
partition function follows as \cite{ali-rudi}
\begin{eqnarray}
\overline{{\mathcal Z}^{n}} = \int \bigg[\prod_{a=1}^n {\mathcal D} \phi_{a}\bigg]\,\rme^{- \beta {\mathcal S}[\phi_{a}(\vct
r)]},
\label{rep-1}
\end{eqnarray}
with
\begin{eqnarray}
{\mathcal S}[\phi_{a}(\vct r)] &=& \frac{1}{2} \sum_{a, b}\int \phi_{a}(\vct r) {\mathcal D}_{a b}(\vct r, \vct r')\phi_{b}(\vct r) \, \rmd\vct r \,\rmd \vct  r' + \rmi \int \bar \rho_0(\vct r)\sum_{a} \phi_{a}(\vct r)\, \rmd\vct r \nonumber\\
& & - \frac{\Lambda_c}{\beta} \int \Omega (\vct r) \sum_{a} \rme^{-
\rmi \beta q e_{0} \phi_{a}(\vct r)} \, \rmd \vct r, 
\label{H-1}
\end{eqnarray}
where $a,b=1,\ldots,n$ are  the replica indices and 
\begin{equation}
{\mathcal D}_{a b}(\vct r, \vct r') =u^{-1}(\vct r, \vct r')\delta_{a
b}
 +\beta g(\vct r) \delta(\vct r - \vct r').
\label{def-1}
\end{equation}

The  expression (\ref{rep-1}) together with Eq. (\ref{free-1}) represents the
starting formulation for the free energy in the presence of quenched charge
disorder. This free energy can only be evaluated approximately. 
Thus, in order to proceed we shall combine the methods developed for the 
one-component (counterions-only) Coulomb fluid without disorder \cite{Netz01} and modify them in order to incorporate appropriately the
disorder effects. We shall start with the WC limit giving rise to the corresponding mean-field theory and then proceed to the SC limit.

\subsection{Disorder effects in the WC regime}

In the WC   limit $\Xi\rightarrow 0$, one may proceed by employing
a saddle-point analysis of the functional integral (\ref{rep-1}), just as in the
case with no disorder in Section \ref{subsec:WCdressed}. It is easy to show that the real-valued mean-field
replica potential $\psi_a$ is governed by the following equation, 
\begin{equation}
-\varepsilon\varepsilon_{0} \nabla^{2} \psi_{a}(\vct r) + \beta
g(\vct r) \sum_{b=1}^n  \psi_{b}(\vct r) = 
 \bar \rho_0(\vct r) + q e_0 \Lambda_c  \Omega(\vct r) \,\rme^{-\beta q e_0 \psi_a}.
\end{equation}
In the replica formulation we have to take the limit $n \rightarrow 0$,
which furthermore implies that $\lim_{n \rightarrow 0} \sum_{b} \psi_{b}(\vct r) \rightarrow
0$.  It is thus evident that in the limit $n \rightarrow 0$, the contributions from
the disorder vanish and, because the index $a$ becomes irrelevant, one
recovers the standard PB equation (\ref{eq:PBeq}). 
Therefore, the quenched charge disorder effects completely vanish
 in the WC limit \cite{ali-rudi,netz-disorder}. 

The above result is a consequence of the mean-field approximation and holds 
in the limit $\Xi\rightarrow 0$ 
even if the system is generalized to contain 
additional ionic species or dielectric discontinuities at the bounding surfaces.  The quenched charge disorder however turns out to play a significant role in dielectrically inhomogeneous systems when  electrostatic field fluctuations are taken into account. It can be shown to lead to an additional attractive or repulsive contribution to the total free energy (depending on the  dielectric mismatch and the salt screening in the system) even when the surfaces are assumed to be {\em net-neutral}   \cite{rudiali, disorder-PRL}.

\subsection{Disorder effects in the SC regime}

The partition function (\ref{rep-1}) can be calculated in the SC limit via a virial expansion up to the first nontrivial 
leading order in powers of the fugacity as noted before. The 
canonical SC free energy of the system then follows from Eq. (\ref{free-1}) by using  a standard Legendre transform \cite{ali-rudi}  as 
\begin{eqnarray}
{\mathcal F}_N &=& \frac{1}{2} \int  \bar \rho_0(\vct r) v(\vct r, \vct r') \bar \rho_0(\vct r') \, \rmd \vct r \, \rmd \vct r' + \frac{1}{2} {\rm Tr} \, g(\vct r) v(\vct r, \vct r')  \nonumber\\
& & - N  k_{\mathrm{B}} T \, \ln \int \Omega (\vct R)\, \rme^{-\beta u(\vct R)}\, \rmd \vct R.
\label{fin-1}
\end{eqnarray}
The first term in Eq. (\ref{fin-1}) is nothing but the direct Coulomb interaction between the mean charge densities $\bar \rho_0(\vct r)$ of the fixed charged surfaces (macroions). The second term is an additive 
contribution from the charge disorder, which becomes important  only in dielectrically inhomogeneous systems \cite{disorder-PRL,rudiali} and will be irrelevant in the present study. The third term, however, embodies the disorder effects on the SC level in the presence of a Coulomb fluid. It depends on the single-particle interaction potential
\begin{equation}
  u(\vct R) = q e_{0} \int 
v(\vct r', \vct R) \bar \rho_0(\vct r') \,\rmd \vct r' - \frac{\beta}{2} (q e_{0})^{2}\int g(\vct r') v^{2}(\vct r', \vct R) \,\rmd \vct r', 
\label{u_final}
\end{equation}
where the second term  comes from the disorder variance $g(\vct r)$ and exhibits a non-trivial dependence on the Coulomb kernel $v(\vct r, \vct r')={1}/({4\pi\varepsilon\varepsilon_0\vert\vct r-\vct r'\vert})$. 

\begin{figure}[t]
\begin{center}
	\begin{minipage}[b]{0.48\textwidth}\begin{center}
		\includegraphics[width=\textwidth]{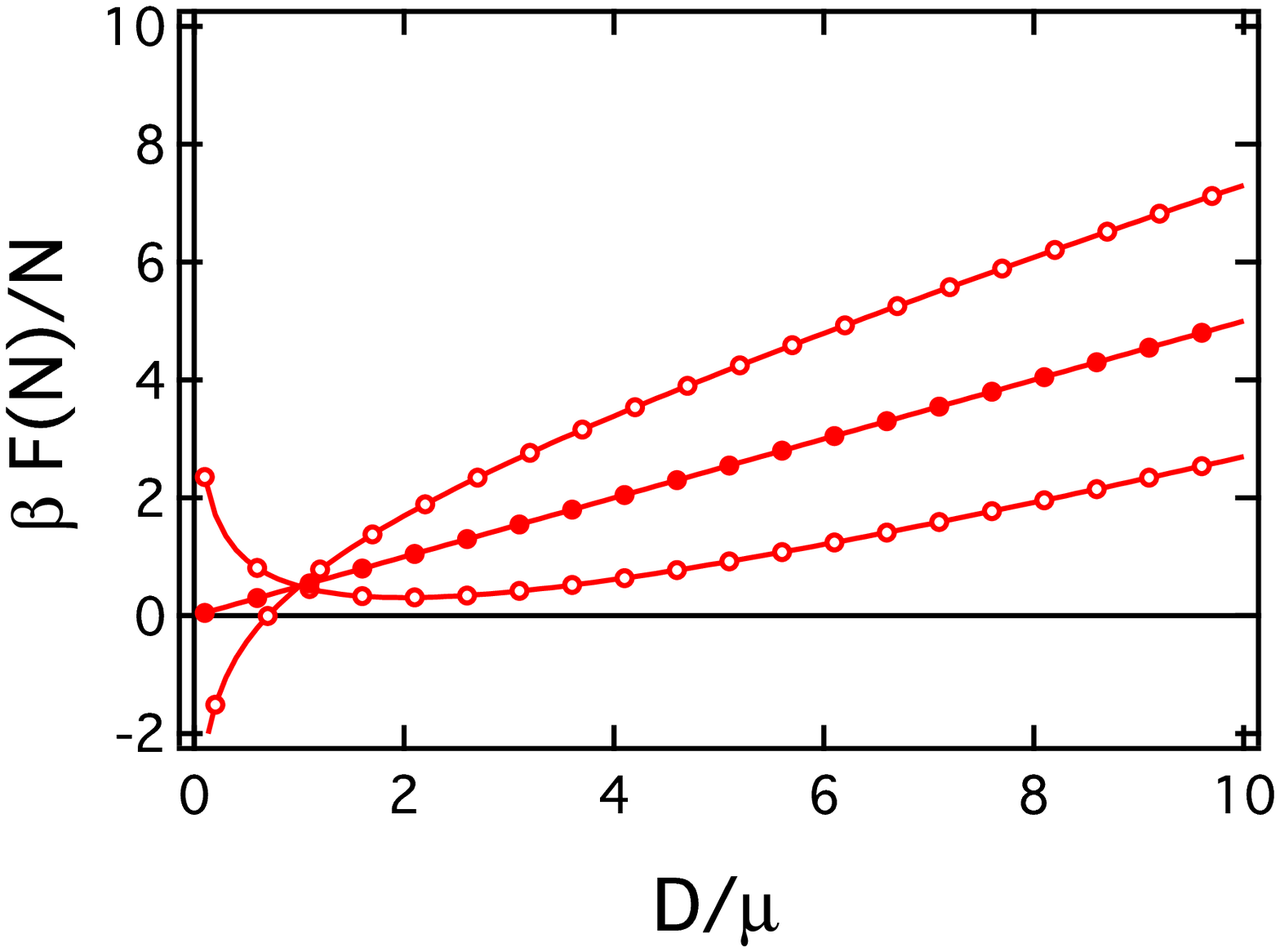} (a)
	\end{center}\end{minipage} \hskip0.25cm
	\begin{minipage}[b]{0.48\textwidth}\begin{center}
		\includegraphics[width=\textwidth]{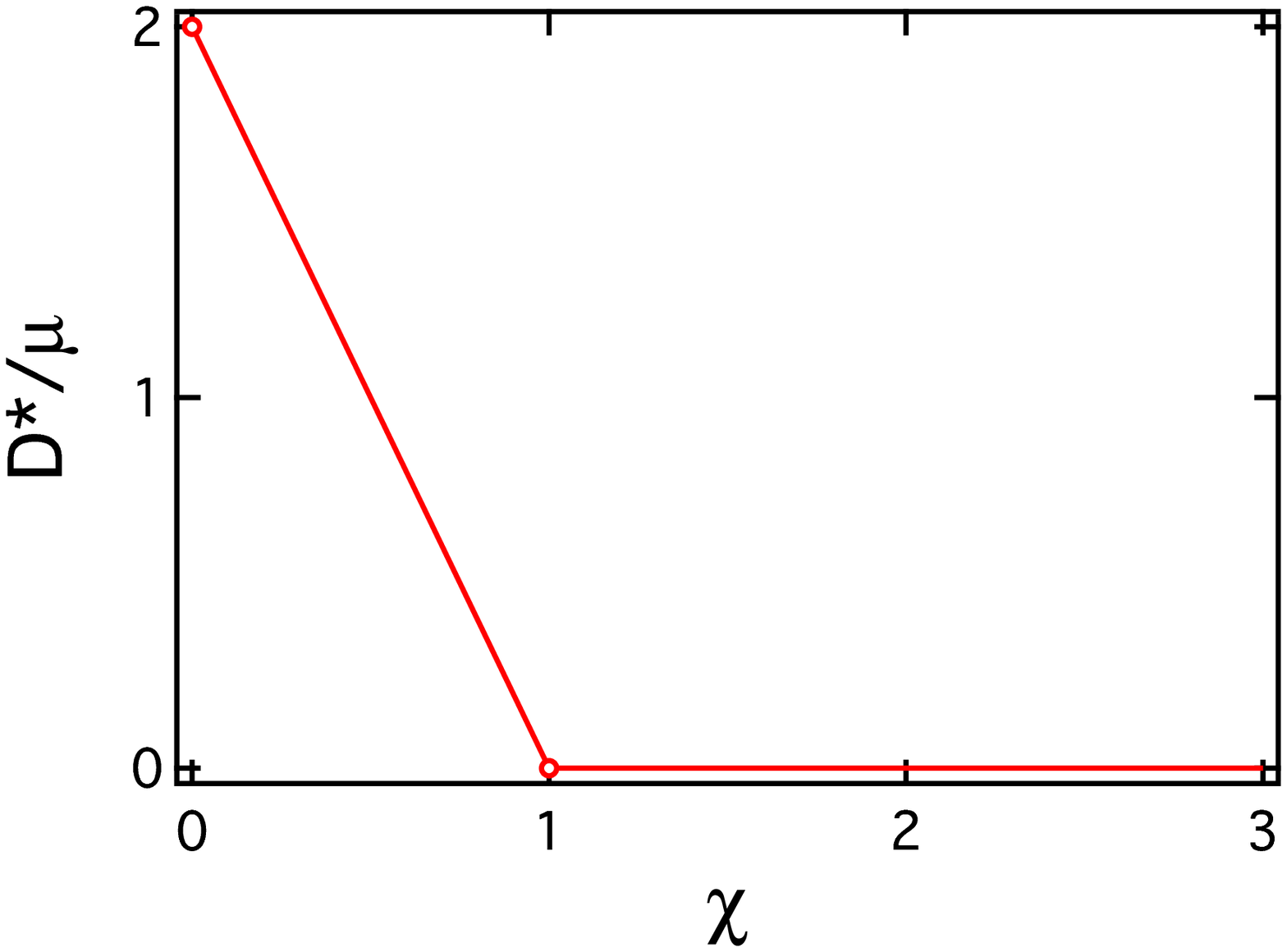} (b)
	\end{center}\end{minipage}
\caption{Left: Rescaled SC free energy, Eq. (\ref{FEN-1}), of two charged
walls bearing quenched charge disorder as a function of
the rescaled inter-surface distance $D/\mu$ for $\chi = 0,1$ and 2 (bottom line, middle line, top line). Right: Rescaled equilibrium distance 
$D^{*}/\mu$ as a function of the disorder coupling parameter $\chi$.}
\label{fig1}
\end{center}
\end{figure}

Assuming again that our system is composed of two planar surfaces located at $z = \pm D/2$ with statistically identical random charge distributions, we write the mean density and variance of the disordered surface charge  as 
\begin{equation}
\bar \rho_0(\vct r) = -\sigma_{\mathrm{s}}e_0 \big[\delta(z - D/2) + \delta(z + D/2) \big] , \quad
g(\vct r) = g e_0^2 \big[ \delta(z - D/2) +  \delta(z + D/2)\big]. 
\end{equation}
The electroneutrality again stipulates that $2  \sigma_{\mathrm{s}} S=  N q $. The geometry function $\Omega(\vct R)$ is the same as before. All the terms in the expression for the free energy, Eq. (\ref{fin-1}), can be computed explicitly. At the end we obtain a surprisingly simple expression
\begin{equation}
\frac{\beta {\mathcal F}_N}{N} 
	=  \frac{D}{2\mu} + (\chi -1)\, \ln\,{D}.
\label{FEN-1}
\end{equation}
Here we have introduced the dimensionless  {\em disorder coupling parameter}
\begin{equation}
\chi = 2\pi q^2\ell_{\mathrm{B}}^2 g, 
\end{equation}
which is very similar to the electrostatic coupling parameter, $\Xi$, in Eq. (\ref{eq:Xi}), except that it is defined based on the disorder variance $g$ and scales with the counterion valency as $q^2$ instead of $q^3$. The free energy  (\ref{FEN-1}) is plotted in Fig. \ref{fig1}a for different values of the disorder coupling parameter.
Note that the disorder leads to a long-range attractive contribution,
which is {\em additive} in the SC free energy  and has a logarithmic 
dependence on the separation, i.e., $\chi \ln \, D$. It thus appears that the quenched charge disorder and the counterions confinement  entropy, i.e., the $- \ln\,{D}$ term in Eq. (\ref{FEN-1}),  in some sense counteract one another.

Evaluating the interaction pressure  from the free energy, Eq. (\ref{FEN-1}), we find $P(D)=  P_{\mathrm{SC}}(D) +  P_{\mathrm{disorder}}(D)$, 
where the first term is the standard SC pressure  \cite{Netz01}, Eq. (\ref{eq:P_SC}),
and the second term is the additive contribution from the disorder 
\begin{equation}
\frac{\beta P_{\mathrm{disorder}}(D)}{2\pi \ell_{\mathrm{B}}\sigma_{\mathrm{s}}^2} =  - {\chi}\bigg(\frac{2\mu}{D}\bigg). 
\label{P-1}
\end{equation}
We can then derive the equilibrium distance $D^{*}$ between
the two surfaces, corresponding to zero interaction pressure, as 
\begin{equation}
D^{*} =  2\,(1 - \chi)\,\mu.
\end{equation}
In the undisordered case, $\chi = 0$, this reduces to the known result
$D^{*} =  2\mu$ \cite{Netz01}, which corresponds to a stable bound state 
for the two surfaces at a separation equal to twice the GC length. 
However, as $\chi$ is increased, 
the equilibrium  
separation decreases and vanishes at the critical value $\chi_c = 1$ and remains at zero thereafter. This behavior has all the features of a second-order, quenched-disorder-induced {\em collapse transition} with an unusual value of the critical exponent (see Fig. \ref{fig1}b). Note also that for $\chi = 1$ the interaction pressure between the surfaces is obviously constant in the whole
range of separations $D$ right down to zero as the
counterions confinement entropy is  completely wiped out by the charge disorder contribution.

%%%%%%%%%%%%%%%%%%%%%%%%%%%%%%%%%%%%%%%%%%%%%%%%%%%%%%%%%%%%%%%%%%%%
\section{Lessons}

The two limiting laws, i.e., the WC and the SC limits for Coulomb fluids that we explored above, are valid in disjoint regions of the parameter space. While the WC limit is valid for sufficiently small macroion surface charge densities, low counterion valencies, high medium dielectric constant and/or high temperatures, the SC limit becomes valid for respectively opposite parameter values. The two together bracket the region of all possible behaviors of Coulomb fluids confined between charged boundaries, a view that was completely corroborated by extensive simulation studies.

The parameter space in between these limiting values can be analyzed by approximate methods \cite{Forsman04,intermediate_regime,Weeks} but is most often accessible solely via computer simulations \cite{hoda_review,Naji_PhysicaA,Netz01,AndreNetz,Naji-cylinders,asim,Weeks,original_sims,Forsman04,trulsson,jho-prl,Naji_CCT, SCdressed,Matej-cyl}.  Exact solutions  for the whole range of coupling parameters are unfortunately available only in one dimension \cite{exact}. The {\em WC-SC paradigm} has been tested extensively \cite{hoda_review,Naji_PhysicaA,Netz01,AndreNetz,Naji-cylinders,asim,Forsman04,intermediate_regime,Weeks,jho-prl,Naji_CCT,exact,Matej-cyl,SCdressed} and fits computer simulations quantitatively correctly in the respective regimes of validity, thus providing a unifying conceptual framework of the behavior of Coulomb fluids.

Though we have shown that in the important limit of SC the much cherished and widespread PB approach does not work, formally its applicability can be systematically extended by perturbative corrections in the local potential fluctuations and correlations \cite{attard,podgornik,fluctuations,david-ron,asim, podgornikparsegianPRL,Netz01,AndreNetz,Netz-orland} along the lines of the standard approach used in the mean-field context \cite{ziherl}. This kind of fix is nevertheless severely limited since the perturbative expansion is only weakly convergent  \cite{Netz01,AndreNetz,hoda_review,Naji_PhysicaA} and  higher-order corrections beyond the first-loop Gaussian term are very complicated and difficult to carry through \cite{david-ron,podgornikparsegianPRL}. Such perturbative corrections offer in effect only a relatively insignificant improvement over the PB approximation \cite{Netz01,AndreNetz,hoda_review,Naji_PhysicaA,asim} and can not  predict phenomena such as like-charge attraction \cite{original_sims}.

The emerging world of Coulomb interactions reviewed above is indeed fascinating. While the counterion-mediated electrostatic interactions between equally charged surfaces are always repulsive on the WC level, the SC regime offers a much richer framework with plethora of new phenomena. Interactions between equally charged macromolecular surfaces can be attractive for strongly coupled counterions, or non-monotonic--showing repulsion at small and large as well as attraction at intermediate separations--for strongly coupled counterions in the presence of weakly coupled simple salt. While charge disorder on macromolecular surfaces has no effect on the WC level, it can quite unexpectedly lead to strong electrostatic attractions between randomly charged
surfaces on the SC level. In view of these developments, the commonly held pop culture wisdom that likes repel and opposites attract should thus be {\em substantially amended!}

%%%%%%%%%%%%%%%%%%%%%%%%%%%%%%%%%%%%%%%%%%%%%%%%%%%%%%%%%%%%%%%%%%%%
\section{Acknowledgment}

R.P. would like to acknowledge the financial support by the Slovenian Research Agency under contract
Nr. P1-0055 (Biophysics of Polymers, Membranes, Gels, Colloids and Cells) and under contract Nr. J1-0908 (Dispersion force nanoactuators). M.K. would like to acknowledge the financial support by the Slovenian Research Agency under the young researcher grant. A.N. is supported by a Newton International Fellowship from the Royal Society, the Royal Academy of Engineering, and the British Academy.

%%%%%%%%%%%%%%%%%%%%%%%%%%%%%%%%%%%%%%%%%%%%%%%%%%%%%%%%%%%%%%%%%%%%

\end{document}